\documentclass{IEEEtran}
\usepackage{amsmath,epsfig} 

\usepackage{amsmath}
\usepackage{amsthm}
\usepackage{amsfonts}
\usepackage{amssymb}
\usepackage{cite}
\usepackage{graphicx}
\usepackage{cases}
\usepackage{cancel}
\usepackage{subfigure}
\usepackage{float}

\def\m(#1){\mathcal{#1}}
\def\bbf(#1){\mathbf{#1}}

\def\row(#1,#2){#1^{\textrm{row}}_#2}
\def\col(#1,#2){#1^{\textrm{col}}_#2}

\newcommand{\rect}{\operatorname{rect} }

\def\colorname(#1){#1}
\newtheorem{theorem}{\colorname(Theorem)}

\newtheorem{proposition}{\colorname(Proposition)}

\newcommand{\RonenComment}[1]{#1} 

\title{Multichannel Sampling of Pulse Streams at the Rate of Innovation}
\author{Kfir Gedalyahu*, Ronen Tur* and Yonina C.~Eldar,~\IEEEmembership{Senior~Member,~IEEE}
\thanks{*Both authors contributed equally to this work.}
\thanks{Department of Electrical Engineering, Technion---Israel Institute of Technology, Haifa 32000, Israel. Phone: +972-4-8293256, fax: +972-4-8295757,
E-mail: \{kfirge@techunix,ronentur@techunix,yonina@ee\}.technion.ac.il.
Y. Eldar is currently a visiting Professor at Stanford, USA.}
}

\begin{document}
\bibliographystyle{IEEEtran}
\maketitle
\begin{abstract}
We consider minimal-rate sampling schemes for infinite streams of delayed
and weighted versions of a known pulse shape.
The minimal sampling rate for these parametric signals is referred to as the rate of
innovation and is equal to the number of degrees of freedom per unit time.
Although sampling of infinite pulse streams was treated in
previous works, either the rate of innovation was not achieved, or
the pulse shape was limited to Diracs.
In this paper we propose a
multichannel architecture for sampling pulse streams with arbitrary shape,
operating at the rate of innovation. Our approach is based on
modulating the input signal with a set of properly chosen
waveforms, followed by a bank of integrators.
This architecture is motivated by recent work on sub-Nyquist sampling of multiband signals.
We show that the
pulse stream can be recovered from the proposed minimal-rate
samples using standard tools taken from spectral estimation in a
stable way even at high rates of innovation. In addition, we
address practical implementation issues, such as reduction of
hardware complexity and immunity to failure in the sampling
channels. The resulting scheme is flexible and exhibits better
noise robustness than previous approaches.
\end{abstract}




%

\section{Introduction}
Digital processing has become ubiquitous, and is the most common
way to process analog signals. Processing analog signals digitally
must be preceded by a sampling stage, carefully
designed to retain the important features of the analog signal
relevant for the processing task at hand. The well known
Shannon-Nyquist theorem states that in order to perfectly
reconstruct an analog signal from its samples, it must be sampled
at the Nyquist rate, i.e., twice its highest frequency. This
assumption is required when the only knowledge on the signal is that
it is bandlimited. Other priors on signal structure
\cite{BeyondBanlimitedSampling_YoninaTomer2009,TomerYoninaSamplingBookChapter},
which include subspace \cite{50yearsAfterShannon,nonideal,Aldroubi}, sparsity
\cite{Yonina2009compressed,mishali2009blind,RobustRecovery_UnionOfSubspaces_YoninaMishali2009}, or smoothness priors
\cite{BeyondBanlimitedSampling_YoninaTomer2009,regular,interpol}, can lead to more
efficient sampling.

An interesting class of structured signals was suggested by Vetterli et al.
\cite{Vetterli2002_Basic,VetterliMagazine2008}, who considered
signals with a finite number of degrees of freedom per unit time,
termed by the authors as signals with \textit{finite rate of
innovation} (FRI). For such models, the goal is to design a
sampling scheme operating at the innovation rate, which is the
minimal possible rate from which perfect recovery is possible. A
special case that was treated in detail are signals consisting of
streams of short pulses. Pulse streams are prevalent in
applications such as bio-imaging \cite{RonenSingleChannel},
neuronal activity and ultra-wideband communications. Since the
pulses are highly compact in time, standard sampling methods
require very high sampling rates. The main idea is to exploit the
fact that the pulse shape is known, in order to characterize such
signals by the time-delays and amplitudes of the various
pulses. Targeting these parameters allows to reduce the sampling
rate way beyond that dictated by the Shannon-Nyquist theorem.
In fact, it was shown in \cite{Ewa} that the sampling rate can also be reduced
for short pulses with unknown shape.

Following this parametric point of view,
a sampling scheme for periodic streams of pulses was developed in
\cite{Vetterli2002_Basic,VetterliMagazine2008}, which operates at the innovation rate.
It relies on the observation that the time delays and amplitudes can be
recovered from a set of the signal's Fourier series coefficients. This follows from the fact
that in the frequency domain, the problem translates into
estimating the frequencies and amplitudes of a sum of complex
sinusoids (cisoids), a problem which has been treated extensively
in the context of spectral estimation \cite{Stoica1997}.

In practical applications finite and infinite streams are usually
encountered, rather than periodic streams.
%
For the finite case Gaussian \cite{Vetterli2002_Basic}, and
polynomials or exponentials reproducing sampling kernels
\cite{DragottiStrangFix2007}, were introduced. The approaches based on the first two kernels, are unstable for high rates of innovations \cite{RonenSingleChannel}.
An alternative sampling scheme, based on a new family of time-limited filters, was presented in
\cite{RonenSingleChannel}. This approach exhibits better noise robustness than previous methods, and is stable even for high model orders.
Exploiting the compact support of the sampling kernels in
\cite{DragottiStrangFix2007,RonenSingleChannel}, both methods were
extended to the infinite case. Unfortunately, neither techniques
achieve the minimal sampling rate, which is the rate of
innovation, for infinite pulse streams.

All previous methods were composed of a single sampling channel.
Multichannel sampling schemes offer additional degrees of freedom
which can be utilized to achieve the rate of innovation for the infinite
setting.
In \cite{Baboulaz} a multichannel extension of the method in \cite{DragottiStrangFix2007} was presented. This scheme allows reduced sampling rate in each channel, but overall sampling rate similar to \cite{DragottiStrangFix2007}.
Another multichannel system, composed of two first-order resistor-capacitor
(RC) networks, was proposed in \cite{UnserFRI2008}. However, this approach assumed that there is a single pulse per sampling period, an assumption which limits the method's time resolution.
Two alternative multichannel methods, were proposed in \cite{kusuma2006multichannel} and \cite{Olkkonen}. These approaches, which are based on a chain of integrators \cite{kusuma2006multichannel} and exponential
filters \cite{Olkkonen}, allow only sampling of infinite streams of Diracs at the rate of innovation.
In addition, we show in simulations that these
methods are unstable, especially for high rates of innovation.
An alternative scheme, proposed in \cite{KfirYonina2009}, can operate at the rate of innovation for pulses with arbitrary shape. However, this approach constrains the delays to be constant in each period.
To the best of our knowledge, a stable minimal-rate sampling
scheme for infinite pulse streams, with arbitrary shape, is still lacking.

Our first contribution treats finite pulse streams. We design a
multichannel sampling system, based on oscillators, mixers and
integrators. In each channel the signal is modulated by an
appropriate waveform, followed by integration over a compact time
interval. We derive conditions which guarantee that the output of
each channel is a mixture of the Fourier coefficients of the
signal. By properly choosing the mixing parameters, we show that
the Fourier coefficients can be obtained from the samples. Once
the set of Fourier coefficients is known, we use standard spectral
estimation tools in order to recover the unknown times and
amplitudes. As we show, the mixing scheme enables simple and
practical generation of modulating waveforms. Furthermore, mixing
the coefficients allows recovering the signal even when one or
more sampling channels fails.

Integration over a finite interval enables a simple extension to the infinite setting.
Our infinite sampling approach leads to perfect reconstruction of the signal, while sampling at
the rate of innovation.
In addition, our scheme can accommodate general pulse shapes with finite length support.
As we show in simulations, our approach exhibits
better noise robustness compared to previous methods, and allows sampling at high
rates of innovation.
We also discuss a special case of infinite streams of pulses having a
shift-invariant (SI) structure, a model presented in \cite{KfirYonina2009}, and
compare our method with the one in \cite{KfirYonina2009}.
Finally, we describe how to practically generate the modulating
waveforms and
derive conditions on these waveforms which guarantee perfect
reconstruction of the signal.

The scheme derived in this work follows ideas of a recently proposed sampling methodology for structured analog signals, termed Xampling \cite{mishali751xampling,xampling_cit},\cite{Ewa}. This framework utilizes the signal model in order to reduce the sampling rate below the Nyquist rate. A pioneer sub-Nyquist system for multiband signals \cite{mishali2009blind}, referred to as the modulated wideband converter (MWC), was proposed in \cite{mishali2010FromTheoryToPractice}. Although treating a different signal model, our modulation scheme is based on concepts presented in \cite{mishali2010FromTheoryToPractice}. Both works share a similar analog front-end, so that the hardware prototype of the MWC, designed in \cite{mishali751xampling}, can also be used to implement our method.



The remainder of this paper is organized as follows. In
Section~\ref{sec_finiteFRI} we derive a multichannel scheme for
finite pulse streams. Section~\ref{sec_infinite} extends our
results to the infinite case. We discuss the generation of the
modulating waveforms in Section~\ref{sec_implementation}, and
present a practical sampling scheme which can be implemented in
hardware. In Section~\ref{sec_related_work} we discuss in more details
the relations of our results to previous work.
Numerical experiments are described in Section~\ref{sec_simulations}.

\section{Finite Streams of Pulses}\label{sec_finiteFRI}
\subsection{Problem Formulation}
Throughout the paper, we denote matrices and vectors by bold font, with lowercase letters corresponding to vectors and uppercase letters to
matrices. The $n$th element of a vector $\mathbf{a}$ is written as $\mathbf{a}_{n}$, and $\mathbf{A}_{ij}$ denotes the
$ij$th element of a matrix $\mathbf{A}$. Superscripts $\left(\cdot\right)^{*}$, $\left(\cdot\right)^{T}$ and
$\left(\cdot\right)^{H}$ represent complex conjugation, transposition and conjugate transposition, respectively. The
Moore-Penrose pseudo-inverse of a matrix $\mathbf{A}$ is written as  $\mathbf{A}^{\dagger}$.
We denote by $\textrm{diag}(\bbf(a))$ a diagonal matrix having the elements of the vector $\bbf(a)$ on its diagonal.
The continuous-time Fourier transform (CTFT) of a
continuous-time signal $x\left(t\right)\in L_{2}$ is defined by
$X\left(\omega\right)=\int_{-\infty}^{\infty}x\left(t\right)e^{-j\omega t}{\rm d}t$.

Consider the finite stream of pulses
\begin{equation}\label{eq_sig_model_aperiodic}
  x(t) = \sum_{l=1}^L a_l h(t - t_l),\quad t_l \in I \subset [0,T),\, a_l \in \mathbb{C},
\end{equation}
where $h(t)$ is a known pulse shape, $\{t_l,a_l\}_{l=1}^L$ are the unknown
delays and amplitudes, and $I$ is a continuous-time interval in $[0,T)$.
The pulse can be arbitrary as long as
\begin{equation}\label{eq_signal_independent_periods_condition}
  h(t - t_l) = 0,\, \forall t\notin [0,T)\quad l=1\ldots L,
\end{equation}
i.e., the signal $x(t)$ is confined to the time-window $[0,T)$.
This condition suggests that the support of the pulse has to be finite and smaller than $T$.
In such cases the effective Nyquist rate will be quite large since $x(t)$ will have a large bandwidth.
However, the sampling rate can be reduced below the Nyquist rate, by noticing that $x(t)$ is uniquely defined by the delays and amplitudes.
Since $x(t)$ has $2L$ degrees of freedom, $\{t_l,a_l\}_{l=1}^L$, at least $2L$ samples are required in order to represent the signal. Our goal is to design a sampling and reconstruction method which perfectly recovers $x(t)$ from this minimal number of samples.

\subsection{Relation to Model-Based Complex Sinusoids Estimation}
Our sampling problem can be related to the well known problem of a model-based complex sinusoids (cisoids) parameter estimation.
This approach was originally taken by Hou and Wu \cite{ziqiang1982nmh}, who were the first to show that time delay estimation can be converted into a frequency estimation of a sum of cisoids \cite{Stoica1997}. This follows from noticing that delays in the time domain are converted into modulations in the frequency domain. However, their method relied on Nyquist rate sampling of the signal, and their derivations were only approximate. Vetterli et al. \cite{Vetterli2002_Basic} addressed this problem from an efficient sampling point of view, and derived a low-rate sampling and reconstruction scheme for periodic streams of Diracs. Their method was based on the same fundamental relation between the delays in time and modulations in frequency.
Following a similar path, we show that once a set of Fourier coefficients of the signal are known,
the delays can be retrieved using sinusoidal estimation methods.
We then design low-rate sampling schemes for obtaining the Fourier coefficients.


Since $x(t)$ is confined to the interval $t\in [0,T)$, it can be expressed by its Fourier series
\begin{equation}
  x(t) = \sum_{k\in\mathbb{Z}} X[k] e^{j\frac{2\pi}{T} k t},\quad t\in [0,T)
\end{equation}
where
\begin{equation}\label{eq_X_k_def}
  X[k] = \frac{1}{T} \int_0^T x(t) e^{-j\frac{2\pi}{T} k t} {\rm d}t.
\end{equation}
Substituting \eqref{eq_sig_model_aperiodic} into \eqref{eq_X_k_def} we obtain
\begin{align}
\nonumber
  X[k] &= \frac{1}{T}\sum_{l=1}^L a_l \int_0^T h(t-t_l) e^{-j\frac{2\pi}{T} k t} {\rm d}t \\
\nonumber
  &= \frac{1}{T}\sum_{l=1}^L a_l e^{-j\frac{2\pi}{T} k t_l} \int_{-\infty}^{\infty} h(t) e^{-j\frac{2\pi}{T} k t} {\rm d}t \\
\label{eq_X_k}
  &= \frac{1}{T}H\left(\frac{2\pi}{T} k\right) \sum_{l=1}^L a_l e^{-j\frac{2\pi}{T} k t_l},
\end{align}
where the second equality stems from the condition in \eqref{eq_signal_independent_periods_condition}, and $H(\omega)$ denotes the CTFT of $h(t)$.

Denote by $\mathcal{K}$ a set of $K$ consecutive indices for which $H\left(\frac{2\pi}{T} k \right) \neq 0,\, \forall k\in\mathcal{K}$.
We require that such a set exists, which is usually the case for short time-support pulses $h(t)$. Denote by $\bbf(H)$ the $K
\times K$ diagonal matrix with \textit kth entry $\frac{1}{T}H\left(\frac{2\pi}{T} k\right)$, and by $\bbf(V)(\bbf(t))$
the $K \times L$ matrix with \textit{kl}th element $e^{-j\frac{2\pi}{T} k t_l}$, where $\bbf(t) = \{t_1,\ldots,t_L\}$ is the
vector of the unknown delays. In addition denote by $\bbf(a)$ the length-$L$ vector whose \textit lth element is $a_l$, and
by $\bbf(x)$ the length-$K$ vector whose \textit kth element is $X[k]$. We may now write \eqref{eq_X_k} in matrix form as
\begin{equation}\label{eq_X_k_matrix}
\bbf(x) = \bbf(H) \bbf(V)(\bbf(t)) \bbf(a).
\end{equation}
The matrix $\bbf(H)$ is invertible by construction, and therefore we can define $\bbf(y) = \bbf(H)^{-1} \bbf(x)$, which satisfies
\begin{equation}\label{eq_y_equal_x_deconvolved}
  \bbf(y) = \bbf(V)(\bbf(t)) \bbf(a).
\end{equation}
Addressing the $k$th element of the vector $\bbf(y)$ in \eqref{eq_y_equal_x_deconvolved} directly, we obtain
\begin{equation}
  \bbf(y)_k = \sum_{l=1}^L a_l e^{-j\frac{2\pi}{T} k t_l}.
\end{equation}

Given the vector $\bbf(x)$, \eqref{eq_y_equal_x_deconvolved} conforms with the standard problem of finding the frequencies and
amplitudes of a sum of $L$ cisoids. The time-delays can be estimated using nonlinear techniques, e.g., the annihilating filter \cite{Vetterli2002_Basic}, matrix-pencil \cite{matrixpencil}, Kumaresan and Tufts method\cite{kumaresan1983eaa} or ESPRIT \cite{ESPRIT_Kailath} (see \cite{Stoica1997} for a review of this topic), as long as $K\geq 2L$ and the time-delays are distinct, i.e., $t_i \neq t_j$ for all $i \neq j$. Once the time-delays are known, the linear set of equations \eqref{eq_y_equal_x_deconvolved} may be solved via least-squares for the unknown amplitudes. Due to the Vandermonde form of $\bbf(V)(\bbf(t))$,
it is left invertible as long as $K \geq L$, so that
$\bbf(a) = \bbf(V)^{\dagger}(\bbf(t)) \bbf(y)$.

\subsection{Direct Multichannel Sampling}
\label{subsec_direct}
As we have seen, given a vector of $K\geq2L$ Fourier series coefficients $\bbf(x)$, we may use standard tools from
spectral analysis to determine the set $\{t_l,a_l\}_{l=1}^L$. In practice, the
signal is sampled in the time-domain, and therefore we do not have direct access to samples of $\bbf(x)$. Our goal now is to design a sampling scheme which will allow to obtain the vector $\bbf(x)$ from time-domain samples.

\begin{figure}[h]
\centering
\includegraphics[scale=0.65]{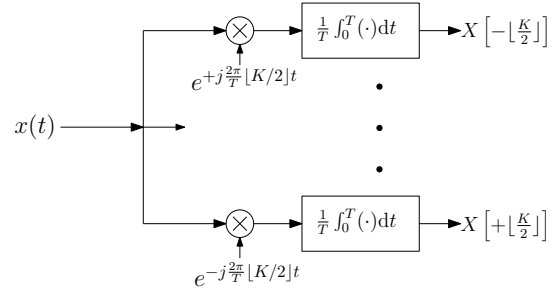}
\caption{Multichannel direct sampling of the Fourier series coefficients $X[k],\, k\in \mathcal{K}$.}
\label{fig_direct_multichannel_X_k_sampling}
\end{figure}

For simplicity, we set $K$ to be an odd number, and choose the set $\mathcal{K} = \{-\lfloor K/2 \rfloor,\ldots,\lfloor K/2 \rfloor\}$. However, our results extend to any set $\mathcal{K}$ of consecutive indices, as long as $|\mathcal{K}| \geq 2L$. The Fourier coefficients $X[k]$ can be obtained using the multichannel sampling scheme depicted in Fig.\ref{fig_direct_multichannel_X_k_sampling}. Each channel consists of modulating $x(t)$ with a complex exponential, followed by an integrator over the window $[0,T)$. The sample taken by the \textit kth channel is exactly $X[k]$, as in \eqref{eq_X_k_def}.
This direct sampling scheme is straightforward, and may be implemented using 3 basic building blocks: oscillators, mixers and integrators. However, from a practical point of view this approach has the disadvantage that it requires many oscillators, having frequencies which must be exact multiples of some common base frequency.


\subsection{Mixing the Fourier Coefficients}\label{subsec_finite_mixing}
We now generalize our framework, towards a more practical sampling scheme by mixing several Fourier coefficients, rather than limiting ourselves to one coefficient per channel. The additional degrees of freedom offered by this extension will allow the design of waveforms that are easy to implement. Our approach is motivated by the hardware reported in \cite{mishali2010FromTheoryToPractice}, where similar modulators are used to sample multiband signals at sub-Nyquist rates.

In addition, in real-life scenarios one or more channels might fail, due to malfunction or noise corruption, and therefore we lose the information stored in that channel.
Unique recovery of the signal parameters from \eqref{eq_X_k_matrix}, relies on having a consecutive set of Fourier coefficients \cite{VetterliMagazine2008}.
Hence, when using the direct scheme, loss of Fourier coefficients, prevents us from recovering the signal. In contrast, when mixing the coefficients we distribute the information about each Fourier coefficient between several sampling channels. Consequently, when one or more channels fail, the required Fourier coefficients may still be recovered from the remaining operating channels. If their number is greater than $2L$, the signal can be still perfectly recovered from the samples. We discuss this feature more thoroughly in Section~\ref{sec_implementation}.

\begin{figure}[h]
\centering
\includegraphics[scale=0.65]{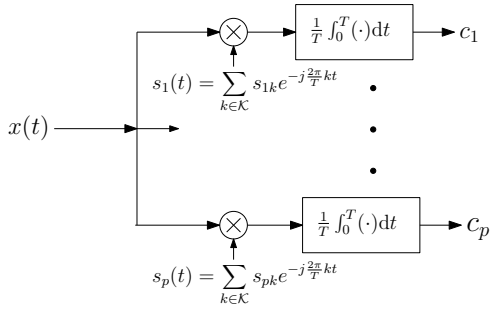}
\caption{Mixing the Fourier coefficients differently in each channel.}
\label{fig_mixing_multichannel_sampling}
\end{figure}

Consider a multichannel sampling scheme with $p$ channels, as depicted in Fig.~\ref{fig_mixing_multichannel_sampling}. In each channel, we modulate the signal using a weighted sum of cisoids given by
\begin{align}
\label{eq:exponent_mix}
s_i(t)=\sum_{k\in\mathcal{K}} s_{ik} e^{-j\frac{2\pi}{T} kt},
\end{align}
where the weights $s_{ik}$ vary from channel to channel. The resulting sample of the \textit ith channel is
\begin{align}
\label{eq_mixing_ith_sample}
  c_i = \frac{1}{T} \int_0^T x(t) \sum_{k\in\mathcal{K}} s_{ik} e^{-j\frac{2\pi}{T} kt} {\rm d}t
    = \sum_{k\in\mathcal{K}} s_{ik} X[k].
\end{align}

To relate the samples and the Fourier coefficients, we define the $p \times K$ matrix $\bbf(S)$ with $s_{ik}$ as its \textit{ik}th element, and by $\bbf(c)$ the length-$p$ sample vector with \textit ith element $c_i$. We may now write \eqref{eq_mixing_ith_sample} in matrix form as
\begin{equation}
\label{eq:mix_eq}
  \bbf(c) = \bbf(S) \bbf(x).
\end{equation}
As long as $\bbf(S)$ has full column rank, where $p\geq K$ is a necessary condition, we can recover $\bbf(x)$ from the samples by $\bbf(x) = \bbf(S)^{\dagger} \bbf(c)$. The direct sampling scheme presented earlier is a special case of this more general approach, with $p=K$ and $\bbf(S) = \bbf(I)$. In Section~\ref{sec_implementation} we exploit the degrees of freedom this general scheme offers, and present sampling schemes which can simplify the hardware design, and are more robust to malfunctions in the sampling channels.

We summarize this result in the following theorem.
\begin{theorem}\label{thm:finite}
  Consider a finite stream of pulses given by
  \begin{equation*}
    x(t) = \sum_{l=1}^L a_l h(t - t_l),\quad t_l \in I \subset[0,T),\, a_l \in \mathbb{C}, \, l=1\ldots L,
  \end{equation*}
  where $h(t)$ is a known pulse shape, and condition \eqref{eq_signal_independent_periods_condition} is satisfied. Choose a set $\mathcal{K}$ of consecutive indices for which $H(2\pi k/T) \neq 0,\, \forall k\in\mathcal{K}$. Consider the multichannel sampling scheme depicted in Fig.~\ref{fig_mixing_multichannel_sampling},
  for some choice of coefficients $\{s_{ik}\}_{k\in\mathcal{K}},\, i=1,\ldots,p$.
  Then, the signal $x(t)$ can be perfectly reconstructed from the samples
  $\{c_i\}_{i=1}^p$ with
  \begin{align}
    c_i &= \frac{1}{T}\int_0^T x(t) \sum_{k\in\mathcal{K}} s_{ik} e^{-j\frac{2\pi}{T} kt} {\rm d}t,
  \end{align}
  as long as $p\geq |\mathcal{K}| \geq 2L$, and the coefficients matrix $\bbf(S)$ in \eqref{eq:mix_eq} is left invertible.
\end{theorem}
As we discuss in Section~\ref{subsec_RelatedWork_SoS}, the method in \cite{RonenSingleChannel} can be viewed as a special case of Fig.~\ref{fig_mixing_multichannel_sampling}. Since our work is a generalization of \cite{RonenSingleChannel}, it benefits from the high noise robustness exhibited by \cite{RonenSingleChannel}, in contrast to previous work \cite{Vetterli2002_Basic,DragottiStrangFix2007}. It should be noted that Theorem~\ref{thm:finite} holds for a periodic pulse stream as well, since it can be similarly represented by a Fourier series, and all derivations remain intact.

We now demonstrate several useful modulating waveforms.
\subsubsection{Cosine and Sine waveforms}
First we set $p=K$. Then, we choose the first $\lfloor K/2 \rfloor$ waveforms to be $\textrm{cos}\left(\frac{2\pi}{T}kt\right)$, the next $\lfloor K/2 \rfloor$ to be $\textrm{sin}\left(\frac{2\pi}{T}kt\right)$, and the last to be the constant function $1$. Clearly, these waveforms fit the form in \eqref{eq:exponent_mix}. It is easily verified that this choice yields an invertible matrix $\mathbf{S}$. The practical advantage of the mixing scheme is already evident, since sine and cosine waves are real valued, whereas the direct multichannel scheme requires complex exponentials.

\subsubsection{Periodic Waveforms}\label{subsubsec_periodic_waveforms}
Every periodic waveform can be expanded into a Fourier series. Transferring such a waveform through some shaping filter, e.g., a low-pass filter, we can reject most of the coefficients, leaving only a finite set intact. Consequently, such a scheme meets the form of \eqref{eq:exponent_mix}. In Section~\ref{sec_implementation} we elaborate on this concept, discuss design considerations, and show that properly chosen periodic waveforms yield a left invertible matrix $\mathbf{S}$.

One simple choice is periodic streams of rectangular pulses modulated by $\pm 1$ \cite{mishali2010FromTheoryToPractice}. The strength of the mixing scheme over the direct one will be emphasized in Section~\ref{subsec_pulses_waveform}. We show that one periodic stream is sufficient for all channels, while each channel uses a delayed version of this common waveform. Therefore, the requirement for multiple oscillators and the need for accurate multiples of the basic frequency, are both removed. In addition, periodic streams are easily designed and implemented digitally, rather than somewhat complicated analog design of oscillators combined with analog circuits intended to create exact frequency multiples. Finally, if the period $T$ changes, the analog circuit has to be modified substantially, whereas the flexibility of the digital design allows simple modifications.

%
%
%

%

\section{Infinite Streams of Pulses}\label{sec_infinite}
\subsection{General Model}
We now consider an infinite stream of pulses defined by
\begin{equation}\label{eq_sig_model_infinite}
  x(t) = \sum_{l\in\mathbb{Z}} a_l h(t - t_l),\quad t_l \in \mathbb{R},\, a_l \in \mathbb{C}.
\end{equation}
We assume that there are no more than $L$ pulses in any interval $I_m \triangleq \left[(m-1)T,mT\right],\, m\in \mathbb{Z}$. We further assume that within each interval condition \eqref{eq_signal_independent_periods_condition} holds, and consequently, the intervals are independent of one another. The maximal number of degrees of freedom per unit time, also known as the rate of innovation \cite{Vetterli2002_Basic,VetterliMagazine2008}, is $2L/T$. We now present a multichannel sampling and reconstruction scheme which operates at the minimal rate possible, i.e., the rate of innovation.

\begin{figure}[h]
\centering
\includegraphics[scale=0.65]{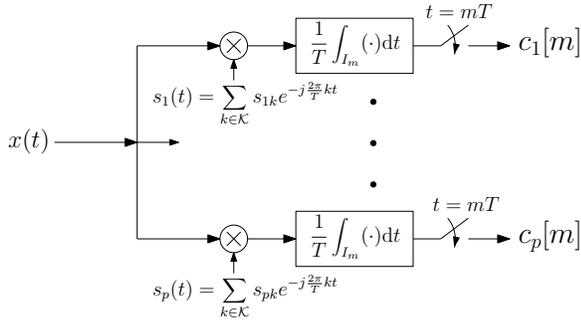}
\caption{Extended sampling scheme using modulating waveforms for an infinite pulse stream.}
\label{fig_infinite_mixing_waveforms_sampling}
\end{figure}

Consider an extension of the sampling scheme presented in Section~\ref{subsec_finite_mixing}, where we sample every $T$ seconds. Upon each sample we reset the integrator, so that the \textit mth sample correspond to an integral over the interval $I_m$. The resulting sampling scheme is depicted in Fig.~\ref{fig_infinite_mixing_waveforms_sampling}.
Since the \textit mth sample is influenced by the interval $I_m$ only, the infinite problem may be reduced into a sequence of finite streams of pulses. The resulting samples are given by
\begin{equation}\label{eq_samples_infinite_matrix_form}
  \bbf(c)[m] = \bbf(S)\bbf(x)[m],
\end{equation}
where the vector $\bbf(x)[m]$ contains the Fourier series coefficients of the signal within the \textit mth interval, $I_m$. As long as $\bbf(S)$ is chosen so that it is left invertible, we can obtain the sequence of Fourier series coefficients by $\bbf(x)[m] = \bbf(S)^{\dagger}\bbf(c)[m]$. Extending \eqref{eq_y_equal_x_deconvolved} to the infinite case we obtain:
\begin{equation}\label{eq_infinite_sum_of_sinusoids_equations}
  \bbf(y)[m] = \bbf(H)^{-1} \bbf(x)[m] = \bbf(V)(\bbf(t)[m]) \bbf(a)[m],
\end{equation}
where $\bbf(t)[m]$ and $\bbf(a)[m]$ are the times and amplitudes of the pulses in the interval $I_m$, respectively, and the matrix $\bbf(V)$ remains as in \eqref{eq_X_k_matrix}. For each $m$, \eqref{eq_infinite_sum_of_sinusoids_equations} represents a sum of cisoids problem, and thus may be solved as long as $K\geq 2L$. By choosing $p=K=2L$ we present a sampling scheme which operates at the rate of innovation, and allows for perfect reconstruction of an infinite stream of pulses.

We state our result in a theorem.
\begin{theorem}\label{thm:infinite}
  Consider an infinite stream of pulses given by
  \begin{equation*}
    x(t) = \sum_{l\in\mathbb{Z}} a_l h(t - t_l),\quad t_l \in \mathbb{R},\, a_l \in \mathbb{C}.
  \end{equation*}
  where $h(t)$ is a known pulse shape. Assume that there are no more than $L$ pulses within any interval $I_m \triangleq \left[(m-1)T,mT\right],\, m\in \mathbb{Z}$, and that condition \eqref{eq_signal_independent_periods_condition} holds for all intervals $I_m,\,m\in\mathbb{Z}$. Choose a set $\mathcal{K}$ of consecutive indices for which $H(2\pi k/T) \neq 0,\, \forall k\in\mathcal{K}$. Consider the multichannel sampling scheme depicted in Fig.~\ref{fig_infinite_mixing_waveforms_sampling},
  for some choice of coefficients $\{s_{ik}\}_{k\in\mathcal{K}},\, i=1,\ldots,p$.
  Then, the signal $x(t)$ can be perfectly reconstructed from the samples
  $\{c_i[m]\}_{i=1}^p$ with
  \begin{align}
  c_i[m] &= \frac{1}{T}\int_{I_m} x(t) \sum_{k\in\mathcal{K}} s_{ik} e^{-j\frac{2\pi}{T} kt} {\rm d}t,\quad m\in\mathbb{Z},
  \end{align}
  as long as $p\geq |\mathcal{K}| \geq 2L$, and the coefficients matrix $\bbf(S)$ in \eqref{eq:mix_eq} is left invertible.
\end{theorem}
To the best of our knowledge Theorem~\ref{thm:infinite} presents the first sampling scheme for pulse streams with arbitrary shape, operating at the rate of innovation. Furthermore, as we show in simulations, our method is more stable than previous approaches.

\subsection{Stream of Pulses with Shift-Invariant Structure}\label{subsec_infinite_SI}
We now focus on a special case of the infinite model \eqref{eq_sig_model_infinite}, proposed in \cite{KfirYonina2009}, where the signal has an additional shift-invariant (SI) structure. This structure is expressed by the fact that in each period $T$, the delays are constant relative to the beginning of the period. Such signals can be described as
\begin{align}\label{eq_sig_model_kfir}
  x(t) = \sum_{m\in\mathbb{Z}}\sum_{\ell=1}^{L} a_{\ell}[m] h(t - t_{\ell}-mT),\, t_{\ell} \in I \subset [0,T),
\end{align}
where $a_{\ell}[n]\in \ell_2$ denotes the $\ell$th pulse amplitude on the $m$th period.
Assuming condition \eqref{eq_signal_independent_periods_condition} holds here as well, \eqref{eq_infinite_sum_of_sinusoids_equations} can be rewritten as
\begin{align}
\label{eq_infinite_eq_for_kfir_model}
  \bbf(y)[m] = \bbf(V)(\bbf(t)) \bbf(a)[m],
\end{align}
since now the relative delays in each period are constant. Here,
$\bbf(a)[m]$ denotes the length-$L$ vector with $\ell$th element $a_{\ell}[m]$.

Clearly, the condition for the general model $p \geq 2L$ is a sufficient condition here also, however the additional prior on the signal's structure can be used to reduce the number of sampling channels.
The results obtained in \cite{KfirYonina2009}, for a similar set of equations, provide the following sufficient condition for unique recovery of the delays and vectors $\bbf(a)[m]$ from \eqref{eq_infinite_eq_for_kfir_model}:
\begin{align}
\label{eq_unique_cond_kfir}
K \geq 2L-\eta+1,
\end{align}
where
\begin{align}
\eta=\textrm{dim}\left(\textrm{span}\left(
\left\{\bbf(a)[m], m\in \mathbb{Z}
\right\}
\right)\right)
\end{align}
denotes the dimension of the minimal subspace containing the vector set $\left\{\bbf(a)[m], m\in \mathbb{Z}\right\}$.
This condition implies that in some cases $K$, and eventually the number of channels $p$ (since $p\geq K$), can be reduced beyond the lower limit $2L$ for the general model, depending on the value of $\eta$.

Similar to \cite{KfirYonina2009}, recovery of the delays from \eqref{eq_infinite_eq_for_kfir_model} can be performed using the ESPRIT \cite{ESPRIT_Kailath} or MUSIC \cite{MUSIC_Schmidt} algorithms. These approaches, known as subspace methods, require that $\eta=L$. In this case they achieve the lower bound of \eqref{eq_unique_cond_kfir}, namely recover the delays using only $p \geq L+1$ sampling channels.
In cases where $\eta<L$, an additional smoothing \cite{smooth} stage is required prior to using the subspace methods, and $p \geq 2L$ sampling channels are needed.

To conclude, when the pulse amplitudes vary sufficiently from period to period, which is expressed by the condition $\eta=L$, the common information about the delays can be utilized to reduce the sampling rate to $(L+1)/T$.
Moreover, the approach presented here can improve the delays estimation in the presence of noise, compared to the one used for the general model, since it uses the mutual information between periods, rather than recovering the delays for each period separately. Will demonstrate this improvement in Section~\ref{subsec_SI_sim}.

This result is summarized in the following theorem.
\begin{theorem}\label{thm:infinite_kfir}
  Consider the setup of Theorem~\ref{thm:infinite}, where now
  \begin{equation*}
    x(t) = \sum_{m\in\mathbb{Z}}\sum_{\ell=1}^{L} a_{\ell}[m] h(t - t_{\ell}-mT),\quad t_l \in I \subset[0,T).
  \end{equation*}
    The signal $x(t)$ can be perfectly reconstructed from the samples $\{c_i[m]\}_{i=1}^p,\,m\in\mathbb{Z}$ as long as the coefficients matrix $\bbf(S)$ in \eqref{eq:mix_eq} is left invertible and
  \begin{equation*}
  p\geq |\mathcal{K}|
  \begin{cases}
     \geq L+1 & \text{when } \eta=L \\
     \geq 2L & \text{when } \eta<L,
  \end{cases}
  \end{equation*}
where $\eta=\textrm{dim}\left(\textrm{span}\left(\left\{\bbf(a)[m], m\in \mathbb{Z}\right\}\right)\right)$
denotes the dimension of the minimal subspace containing the vector set $\left\{\bbf(a)[m], m\in \mathbb{Z}\right\}$.
\end{theorem}

\subsection{Channel Synchronization}
\label{subsec:sync}
The sampling scheme of Fig.~\ref{fig_infinite_mixing_waveforms_sampling} has two main disadvantages relative to single channel-based schemes: each sampling channel requires additional hardware components, and precise synchronization of the channel's sampling times is required.
In this subsection we treat the synchronization issue, and discuss the approaches to overcome it.

One way to synchronize the channels, is on the hardware level, for example by using a zero-delay synchronization device \cite{zero_delay,mishali2010FromTheoryToPractice}.
Such a device produces accurate trigger signals for the samplers and integrators in all the channels.
An alternative approach is to perform a prior calibration process, in which the relative delay of each channel is measured. The calibration can be performed at the system manufacturing stage or during its power-on, by stimulating the system with a known signal.
As we now show, once the time offsets between the channels are known, they can be compensated.

Suppose that the $i$th channel has a time offset of $\Delta_i \in [-\Delta_{\textrm{max}},\Delta_{\textrm{max}}]$
relative to the optimal sampling instants $t=nT$, where $\Delta_{\textrm{max}}$ is the maximal possible offset.
We assume that for each time interval $T$
\begin{align}\label{eq_signal_independent_periods_condition_sync}
  h(t - t_l) = 0,\, \forall t\notin [\Delta_{\textrm{max}},T-\Delta_{\textrm{max}})\quad l=1\ldots L.
\end{align}
This condition ensures that the intervals can be processed independently using the approach we now propose.

The effective delay of the $l$th pulse measured in the $i$th channel is $t_l-\Delta_i$.
Substituting into \eqref{eq_X_k}, the $k$th Fourier coefficient measured in the $i$th channel satisfies
\begin{align}
\label{eq_X_i_k_sync}
  X_i[k] = \frac{1}{T}H\left(\frac{2\pi}{T} k\right)
  \sum_{l=1}^L a_l e^{j\frac{2\pi}{T} k (t_l-\Delta_i)}
   = e^{-j\frac{2\pi}{T} k \Delta_i} X[k].
\end{align}
Therefore, from \eqref{eq_mixing_ith_sample},
\begin{align}
\label{eq:c_i_mix_again2}
c_i = \sum_{k\in\mathcal{K}} s_{ik} e^{-j\frac{2\pi}{T} k \Delta_i} X[k]
= \sum_{k\in\mathcal{K}} \tilde{s}_{ik} X[k],
\end{align}
where we defined
\begin{align}
\label{eq_s_ik_tilde}
\tilde{s}_{ik}=s_{ik} e^{-j\frac{2\pi}{T} k \Delta_i}.
\end{align}

From \eqref{eq_s_ik_tilde} we conclude that when the time offsets between the channels are known, the effective mixing matrix of the system is $\tilde{\bbf(S)}$, a matrix whose $ik$th element is $\tilde{s}_{ik}$.
The misalignment between the sampling channels can be compensated by inverting $\tilde{\bbf(S)}$.
We will further discuss the effects of channel misalignment, with unknown delays, in Section~\ref{subsec_sync_sim}.

\section{Modulation Waveforms}
\label{sec_implementation}
We now thoroughly treat the example in Section~\ref{subsec_finite_mixing} of generating the modulation waveforms using periodic signals.
We first address the case of general periodic waveforms, and then focus on the special case of pulse sequences.

\subsection{General Periodic Waveforms}
Our aim is to show how to obtain the required modulating waveforms \eqref{eq:exponent_mix} using a set of $p$ periodic functions, given by $p_i(t)$.
Such waveforms can be expressed using their Fourier series expansion as
\begin{align}
\label{eq:pi_fourier}
p_i(t)=\sum_{k \in \mathbb{Z}} d_{i}[k] e^{j\frac{2\pi}{T} k t},
\end{align}
where the $k$th Fourier series coefficient of $p_i(t)$ is given by
\begin{align}
d_{i}[k]=\frac{1}{T} \int_{0}^{T} p_i(t) e^{-j\frac{2\pi}{T} k t} dt.
\end{align}

The sum in \eqref{eq:pi_fourier} is generally infinite, in
contrast to the finite sum in \eqref{eq:exponent_mix}. Therefore,
we propose filtering $p_i(t)$ with a filter $g(t)$ which rejects
the unwanted elements in the sum \eqref{eq:pi_fourier}. The
filtered waveforms at the output of $g(t)$ are given by
\begin{align}
\tilde{p}_i(t)=p_i(t) * g(t),
\end{align}
and are also periodic. Therefore they can be written as
\begin{align}
\label{eq:pi_tilde_fourier}
\tilde{p}_i(t)=\sum_{k \in \mathbb{Z}} \tilde{d}_{i}[k] e^{j\frac{2\pi}{T} k t},
\end{align}
where it can be easily verified that
\begin{align}
\label{eq:di_tilde} \tilde{d}_{i}[k]={d}_i[k] \cdot
G\left(\frac{2\pi}{T}k\right).
\end{align}
Here $G(\omega)$ denotes the CTFT of $g(t)$. From
\eqref{eq:di_tilde}, the shaping filter $g(t)$ has to satisfy
\begin{align}
\label{eq:filter_cond}
G(\omega)=
\begin{cases}
\text{nonzero} & \omega = \frac{2\pi}{T}k, \ k \in \mathcal{K} \\
0 & \omega = \frac{2\pi}{T}k, \ k \notin \mathcal{K} \\
\text{arbitrary} & \text{elsewhere},
\end{cases}
\end{align}
so that $\tilde{d}_{i}[k]=0$ for $k \notin \mathcal{K}$. This
condition is similar to that obtained in
\cite{RonenSingleChannel} for single channel sampling. Therefore,
the class of filters developed there, can also be used here as a
shaping filter.

Note that \eqref{eq:filter_cond} implies that the frequency response of $g(t)$ is specified only on the set of discrete points $\frac{2\pi}{T}k, \ k \in \mathbb{Z}$, offering large freedom when designing a practical analog filter. For instance, when implementing a lowpass filter (LPF) this allows a smooth transition band between the passband and the stopband of the filter, with a width of $\frac{2\pi}{T}$.

\begin{figure}[h]
\centering
\includegraphics[scale=0.65]{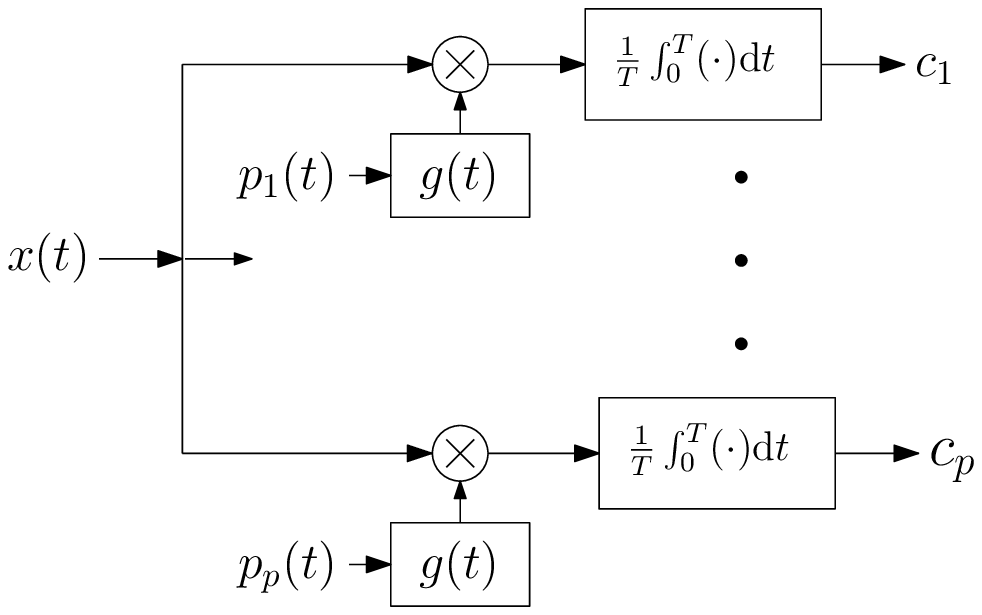}
\caption{Proposed sampling scheme, using modulating waveforms.}
\label{fig_mixing_waveforms_sampling}
\end{figure}

The resulting scheme is depicted in Fig.~\ref{fig_mixing_waveforms_sampling}.
The corresponding elements of the mixing matrix $\bbf(S)$ are given by
\begin{align}
\label{eq:Sil}
\bbf(S)_{ik}=\tilde{d}_{i}[-(k-\lfloor K/2 \rfloor)].
\end{align}
The invertibility of $\bbf(S)$ can be ensured by proper selection of the periodic waveforms $p_i(t)$. In the next subsection we discuss one special case, which allows simple design $\bbf(S)$.

\subsection{Pulse Sequence Modulation}
\label{subsec_pulses_waveform}
We follow practical modulation implementation ideas presented in \cite{mishali2010FromTheoryToPractice,mishali751xampling}, and consider the set of waveforms
\begin{align}
\label{eq:mod_wave}
p_i(t)=\sum_{m \in  \mathbb{Z}} \sum_{n=0}^{N-1}\alpha_i[n] p(t-nT/N-mT)
\end{align}
for $ i=1,\ldots,p$,
where $p(t)$ is some pulse shape and $\alpha_i[n]$ is a length-$N$
sequence. Our aim is to calculate the mixing matrix $\bbf(S)$, when
using the filtered version of \eqref{eq:mod_wave} as modulating
waveforms. To this end, we first compute the Fourier series
coefficients $d_i[k]$ of $p_i(t)$ as
\begin{align}
\label{eq:di_pulses}
{d}_{i}[k]
= & \frac{1}{T} \sum_{m \in  \mathbb{Z}} \sum_{n=0}^{N-1}\alpha_i[n] \int_{0}^{T} p(t-nT/N-mT) e^{-j\frac{2\pi}{T} k t}dt \nonumber \\
= & \frac{1}{T} \sum_{n=0}^{N-1}\alpha_i[n] \sum_{m \in  \mathbb{Z}} \int_{-mT}^{-(m-1)T} p(t-nT/N) e^{-j\frac{2\pi}{T} k t}dt \nonumber \\
= & \frac{1}{T} \sum_{n=0}^{N-1}\alpha_i[n] \int_{-\infty}^{\infty} p(t-nT/N) e^{-j\frac{2\pi}{T} k t}dt \nonumber \\
= & \frac{1}{T} \sum_{n=0}^{N-1}\alpha_i[n] P\left(\frac{2\pi}{T} k \right) e^{-j\frac{2\pi}{N}k n},
\end{align}
where $P(\omega)$ denotes the CTFT of $p(t)$.

 Combining \eqref{eq:di_pulses} with \eqref{eq:Sil} and \eqref{eq:di_tilde},
\begin{align}
\bbf(S)_{ik}=\frac{1}{T} \sum_{n=0}^{N-1}\alpha_i[n]
P\left(\frac{2\pi}{T} k' \right) G \left(\frac{2\pi}{T} k' \right) e^{-j\frac{2\pi}{N}k' n},
\end{align}
where we defined $k'=-(k-\lfloor K/2 \rfloor)$.
The resulting matrix $\bbf(S)$ can be decomposed as
\begin{align}
\label{eq:M_decomp}
\bbf(S)=\bbf(A)\bbf(W)\bbf(\Phi),
\end{align}
where $\bbf(A)$ is a $p \times N$ matrix with $in$th element equal to $\alpha_i[n]$,
$\bbf(W)$ is an $N \times K$ matrix with $nk$th element equal to $e^{-j\frac{2\pi}{N}k'n}$, and
$\bbf(\Phi)$ is a $K \times K$ diagonal matrix with $k$th diagonal element
\begin{align}
\bbf(\Phi)_{k k}=\frac{1}{T} P\left(\frac{2\pi}{T} k' \right) G \left(\frac{2\pi}{T} k' \right).
\end{align}
From this decomposition it is clear that $\bbf(\Phi)$ has to be invertible and the matrix $\bbf(A)\bbf(W)$
has to be left invertible, in order to guarantee
the left invertibility of $\bbf(S)$. We now examine each one of these matrices.

We start with the matrix $\bbf(\Phi)$. From \eqref{eq:filter_cond}, $G\left(\frac{2\pi}{T} k \right) \neq 0$ for $k \in \mathcal{K}$.
Therefore, we only need to require that $P\left(\frac{2\pi}{T} k \right) \neq 0$ for $k \in \mathcal{K}$ in order for $\bbf(\Phi)$ to be invertible.
A necessary condition for the matrix $\bbf(A)\bbf(W)$ to be left invertible, is that $\bbf(W)$ has full column rank.
The matrix $\bbf(W)$ is a Vandermonde matrix, and therefore has full column rank as long as $N \geq K$ \cite{hoffman1971linear}.
The left invertibility of the $p \times K$ matrix $\bbf(A)\bbf(W)$ can be ensured, by proper selection of the sequences $\alpha_i[n]$, where a necessary condition is that $p\geq K$.

We summarize our results in the following proposition:
\begin{proposition}\label{prop_reconstruction_conditions}
Consider the system depicted in Fig.~\ref{fig_mixing_waveforms_sampling}, where the modulation waveforms are given by \eqref{eq:mod_wave}.
If the following conditions hold
\begin{enumerate}
 \item $p \geq K$, $N \geq K$,
 \item The frequency response of the shaping pulse $g(t)$ satisfies \eqref{eq:filter_cond},
 \item The frequency response of the pulse $p(t)$ satisfies $P\left(\frac{2\pi}{T} k \right) \neq 0$ for $k\in\mathcal{K}$,
 \item The sequences $\alpha_i[n]$ are chosen such that the matrix $\bbf(AW)$ has a full column rank,
\end{enumerate}
then the mixing matrix $\bbf(S)$ in \eqref{eq:mix_eq} is left invertible.
\end{proposition}

We now give two useful configurations, that satisfy the conditions of Proposition~\ref{prop_reconstruction_conditions}.

\subsubsection{Single Generator}
We create the $i$th sequence, by taking a cyclic shift of one common sequence $\alpha[n]$ as
\begin{align}
\alpha_i[n]=\alpha[n-i+1 \textrm{ mod } N],
\end{align}
where we assume $p=N$.
Clearly, the corresponding waveforms can be created by using only one pulse generator, where the waveform at the $i$th channel is a delayed version of the generator output, delayed by $(i-1)T/N$ time units.
This suggests, that in contrast to the direct scheme in Fig.~\ref{fig_direct_multichannel_X_k_sampling}, which requires multiple frequency sources, here only one pulse generator is required which simplifies the hardware design.
It is easy to see that with this choice, $\bbf(A)$ will be a circulant matrix.
Such a matrix can be decomposed \cite{MatrixComputationsGolub} as
\begin{align}
\bbf(A)=\bbf(F)^H \textrm{diag} \left(\bbf(F)\boldsymbol{\alpha}\right) \bbf(F),
\end{align}
where $\bbf(F)$ is a $N \times N$ unitary discrete Fourier transform (DFT) matrix, and $\boldsymbol{\alpha}$ is a length-$N$ vector containing the elements of the sequence $\alpha[n]$.
Therefore, for $\bbf(A)$ to be invertible the DFT of the sequence $\alpha[n]$ can not take on the value zero.

\begin{figure}[h]
\centering
\includegraphics[]{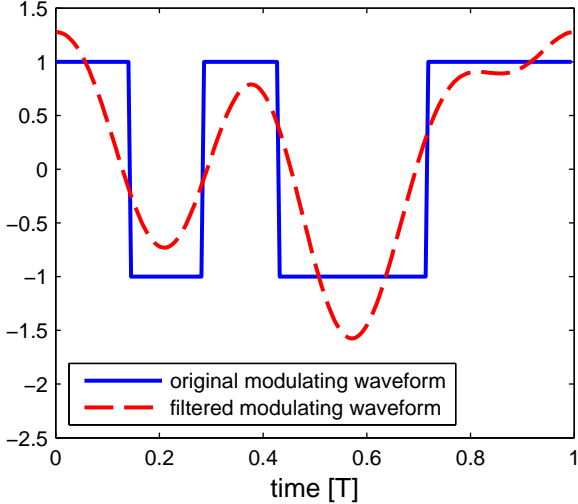}
\caption{Modulating waveform in the time domain, before and after filtering.}
\label{fig_waveform_time}
\end{figure}

\begin{figure}[h]
\centering
\includegraphics[]{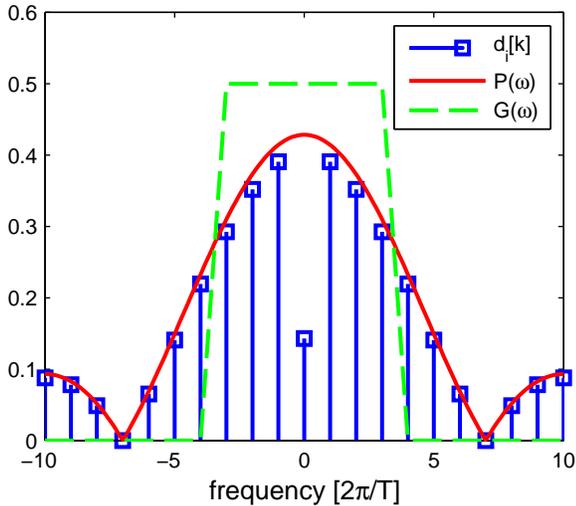}
\caption{Modulating waveform in the frequency domain.}
\label{fig_waveform_freq}
\end{figure}

We now give an example for such a selection of the system's parameters.
We set $p=N=K$, and choose
\begin{align}
p(t)=
\begin{cases}
1 & t\in \left[0,\frac{T}{N}\right] \\
0 & t\notin \left[0,\frac{T}{N}\right].
\end{cases}
\end{align}
The frequency response of this pulse satisfies
\begin{align}
P(\omega)=\frac{T}{N} e^{-j\frac{T}{2N} \omega} \cdot \text{sinc} \left(\frac{T}{2\pi N} \omega\right).
\end{align}
Therefore,
\begin{align}
\left|P\left(\frac{2\pi}{T} k \right)\right|=\frac{T}{N} \text{sinc} \left(\frac{k}{N}\right),
\end{align}
which is non-zero for $k \in \mathcal{K}$.
In addition we choose the sequences $\alpha_i[n]$ as sequences of $\pm 1$s, created from cyclic shifts of one basic sequence, in a way that yields an invertible matrix $\bbf(A)$. Such rectangular pulses with alternating signs can be easily implemented in hardware \cite{mishali751xampling}. In Figs.~\ref{fig_waveform_time} and \ref{fig_waveform_freq}, one modulating waveform is shown in the time and frequency domains, for $p=N=K=7$. The original time-domain waveform is comprised of rectangular pulses, whereas lowpass filtering results in a smooth modulating waveform. Switching to the frequency domain, the Fourier series coefficients are shaped by $P(\omega)$, the CTFT of the pulse shape. The shaping filter frequency response, $G(\omega)$, is designed to transfer only the Fourier coefficients whose index is a member of the set $\mathcal{K}=\{-3,\ldots,3\}$, suppressing all other coefficients.

\subsubsection{Robustness to Sampling Channels Failure}
Next we provide a setup which can overcome failures in a given number of the sampling channels. The identification of the malfunctioning channels is assumed to be performed by some external hardware.

We consider a maximal number $p_e$ of malfunctioning channels and assume that $p\geq N+p_e$.
In order to ensure unique recovery of $\bbf(x)$ from the remaining channels, the submatrix $\tilde{\bbf(S)}$, obtained from $\bbf(S)$ by omitting of the corresponding rows, should be left invertible. Since this has to be satisfied for every possible selection of $p_e$ rows, we need to design $\bbf(S)$ such that any $p-p_e$ rows will form a rank-$K$ matrix.
Following our ideas from the previous discussion, we demonstrate how to reduce the number of required generators, for the current setting.
For simplicity, we assume $p \geq 2p_e$ and that two different generators are used. The first half of the sampling channels use delayed versions of the first generator output, and the second half uses the second generator. By proper selection of the two sequences, the condition mentioned above can be satisfied.

\begin{figure}[h]
\centering
\includegraphics[]{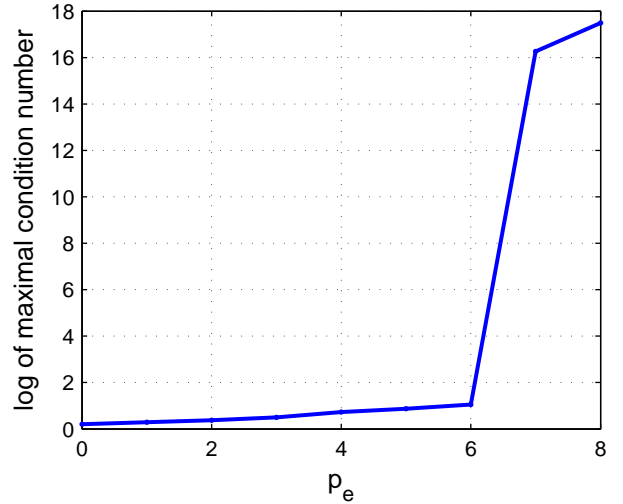}
\caption{Robustness to sampling channels failure example, $N=K=9$, $p=18$, two pulse generators.}
\label{fig_error_correct}
\end{figure}

We now give a numerical example for a such choice. We assume $L=4$ and set $N=K=9$ and $p=18$ sampling channels, which are based on two generators only. Each generator produces a different sequence of $\pm 1$, chosen randomly.
In Fig.~\ref{fig_error_correct} we plot the log of the maximal condition number of $\tilde{\bbf(S)}$, obtained when going over all possible options for omitting $p_e$ rows from $\bbf(S)$.
It can be seen that for $p_e \leq 6$ a relatively low condition number of the matrix $\tilde{\bbf(S)}$ is achieved in the worst case, suggesting that its rank is $N$ as required and $\tilde{\bbf(S)}^{\dag}$ is not ill-conditioned. Therefore, we can overcome failure in up to $6$ sampling channels, using this system. In this case the required $K=9$ consecutive Fourier coefficients can be obtained from the remaining channels, allowing the perfect recovery of the $L=4$ pulses.
In contrast, when channels fail in the direct scheme of Fig.~\ref{fig_direct_multichannel_X_k_sampling}, a set of
$K = 9$ consecutive Fourier coefficients cannot always be obtained. Therefore, perfect recovery of the signal using methods such as annihilating filter or matrix pencil is not guaranteed.

\section{Related Work}\label{sec_related_work}
\subsection{Single-Channel Sampling with the SoS Filter}\label{subsec_RelatedWork_SoS}
The work in \cite{RonenSingleChannel} considered single-channel sampling schemes for pulse streams, based on a filter which is comprised of a Sum of Sincs (SoS) in the frequency domain.
This filter can be expressed in the time domain as
\begin{equation}
  g(t) = \rect\left(\frac{t}{T}\right) \sum_{k \in \mathcal{K}} b_k e^{j\frac{2\pi}{T} kt},
\end{equation}
where $\mathcal{K}$ is the chosen index set, and the coefficients $\{b_k\}_{k\in\mathcal{K}}$ have arbitrary nonzero values.

To explore the relation of our method to \cite{RonenSingleChannel}, we first focus on periodic streams of pulses with period $T$.
By sampling a periodic stream of pulses using the scheme depicted in Fig.~\ref{fig_single_channel_scheme}, the following samples are obtained
\begin{equation}\label{eq_single_channel_samples}
  c[n] = \sum_{k\in\mathcal{K}} b_k X[k] e^{j \frac{2\pi}{T} k n T_s},\quad n=0,\ldots,p-1,
\end{equation}
where $T_s=T/p$ is the sampling period.
Using the matrix $\bbf(V)$ defined in \eqref{eq_X_k_matrix} only now with parameter $\bbf(t)_s = \{0,T_s,\ldots,(p-1)T_s\}$, and defining the diagonal matrix $\bbf(B)$ with $k$th diagonal element $b_k$, \eqref{eq_single_channel_samples} can be written in matrix form as
\begin{equation}
  \bbf(c) = \bbf(V)(-\bbf(t)_s) \bbf(B) \bbf(x).
\end{equation}
Therefore, this is a special case of our multichannel sampling scheme in \eqref{eq:mix_eq} with mixing matrix $\bbf(S) = \bbf(V)(-\bbf(t)_s) \bbf(B)$. The matrix $\bbf(B)$ is invertible by construction, and $\bbf(V)(-\bbf(t)_s)$ is a Vandermonde matrix with distinct times, so that $\bbf(S)$ is left-invertible as long as $p \geq K$. Using this choice of $\bbf(S)$, the $p$ samples taken over one period in \cite{RonenSingleChannel}, are equal to the samples at the output of the $p$ channels in our scheme.
\RonenComment{
\begin{figure}[h]
\centering
\includegraphics[scale=0.8]{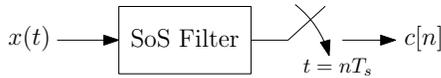}
\caption{Single-channel sampling scheme using the SoS filter.} \label{fig_single_channel_scheme}
\end{figure}}



Exploiting the compact support of the SoS filter, the method was extended to the finite and infinite settings as well \cite{RonenSingleChannel}. The extension is based on using an $r$-fold periodic continuation of the SoS filter $g(t)$, where the parameter $r$ depends on the support of the pulse-shape $h(t)$.
For the infinite case, the samples of the two schemes still coincide. However, since \cite{RonenSingleChannel} is a filtering based scheme, proper separation of at least $1.5T$ between periods is required  in order to obtain independent processing of each period.
While keeping the same sampling rate of $2L/T$, the model in \cite{RonenSingleChannel} has a rate of innovation of $2L/(2.5T)$. Hence the single channel configuration of \cite{RonenSingleChannel} does not achieve the rate of innovation in the infinite case.
%

We now examine the modulation waveforms that result from \cite{RonenSingleChannel}.
Since $\bbf(S) = \bbf(V)(-\bbf(t)_s) \bbf(B)$,
\begin{align}
s_i(t) & =\sum_{k\in\mathcal{K}} b_k e^{-j \frac{2 \pi} {T} k (t-iT/p)}.
\end{align}
It is easily shown that these waveforms can be expressed as
\begin{align}
s_i(t) & = \tilde{g}(-(t-iT/p)),
\end{align}
where $\tilde{g}(t)$ is the periodic continuation of the SoS filter $g(t)$.
Therefore, in each channel the signal is modulated by a delayed version of the periodic SoS filter.
The equivalence of the schemes is easy to explain: sampling the convolution between the input signal and the SoS filter in \cite{RonenSingleChannel}, is equivalent to performing inner products (multiplication followed by integration) with delayed and reflected versions of this filter.
This relation provides another valid class of modulation waveforms.

\subsection{Multichannel Schemes for Shift-Invariant Pulse Streams}

\begin{figure}[h]
\centering
\includegraphics[scale=0.55]{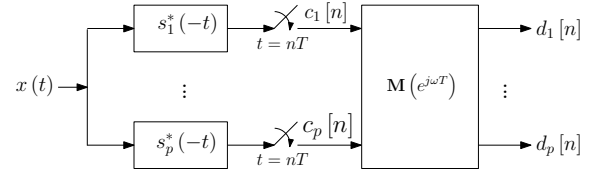}
\caption{Proposed sampling scheme in \cite{KfirYonina2009}.}
\label{fig_kfir_sampling}
\end{figure}

Another related work is \cite{KfirYonina2009} which treats the SI signal model \eqref{eq_sig_model_kfir} presented in Section~\ref{subsec_infinite_SI}.
The sampling scheme proposed in \cite{KfirYonina2009} is depicted
in Fig.~\ref{fig_kfir_sampling}. In each channel, the input signal
is filtered by a band-limited sampling kernel $s_{\ell}^*(-t)$
followed by a uniform sampler operating at a rate of $1/T$. After
sampling, a properly designed digital filter
correction bank, whose frequency response in the DTFT domain is
denoted here by $\bbf(M)(e^{\j\omega T})$, is applied on the
sampling sequences. The exact form of this filter bank is detailed
in \cite{KfirYonina2009}. It was shown in \cite{KfirYonina2009},
that the ESPRIT algorithm can be applied on the corrected
samples, in order to recover the unknown delays.

The sampling rate achieved by the method in \cite{KfirYonina2009}
is generally $2L/T$, where for certain signals it can be reduced
to $(L+1)/T$. Such signals satisfy
$\textrm{dim}(\textrm{span}(\left\{\bbf(d)[m], m\in
\mathbb{Z}\right\}))=L$, where the vectors $\bbf(d)[m]$ contain the samples at the output of the scheme depicted in Fig.~\ref{fig_kfir_sampling}.
This condition is different
than \eqref{eq_unique_cond_kfir}, which directly depends on the vectors
$\bbf(a)[m]$.
Therefore the sampling rate, when using the scheme in \cite{KfirYonina2009}, can
be reduced to $(L+1)/T$ for different signals. This fact
is not surprising, since each approach has a different analog
sampling stage. In both methods, the worst-case minimal sampling
rate is $2L/T$.

The approach in \cite{KfirYonina2009} has two main advantages over the proposed method in this work.
The first is that condition \eqref{eq_signal_independent_periods_condition} is not required. Therefore \cite{KfirYonina2009} can also treat pulses with infinite time support, in contrast to our method.
Another advantage is that it can support single channel configurations.
It was shown in \cite{KfirYonina2009,bajwa_gedal}, that one sampling channel followed by a serial to parallel converter, can be used in order to produce the parallel sampling sequences in Fig.~\ref{fig_kfir_sampling}.

On the other hand, the method depicted in Fig.~\ref{fig_infinite_mixing_waveforms_sampling}
has several advantages over \cite{KfirYonina2009}. First, the equivalent stage for the digital correction in \cite{KfirYonina2009}, is replaced by inversion of the matrices $\bbf(H)$ and $\bbf(S)$, since
\begin{align}
  \bbf(y)[m] = (\bbf(S)\bbf(H))^{-1} \bbf(c)[m].
\end{align}
This operation can be viewed as a one-tap digital correction filter bank, in contrast to the filter $\bbf(M)(e^{\j\omega T})$, which generally has a larger number of taps. Therefore, the proposed correction stage, is much simpler and requires lower computational complexity, than the one in \cite{KfirYonina2009}.

An additional advantage of our scheme, is that the approach of \cite{KfirYonina2009} requires collection of an infinite number of samples, even when the input signal contains a finite number of periods. This requirement is due to the infinite time support of the band-limited sampling kernels.
Moreover, if one is interested only in a finite time interval of the signal, the method in \cite{KfirYonina2009} does not allow processing it separately.
This is in contrast to the proposed scheme, which integrates finite time intervals, and can collect samples only from the relevant periods.
We will demonstrate this advantage in Section~\ref{sec_simulations}.

%

\subsection{Modulated Wideband Converter}
The concept of using modulation waveforms, is based on ideas which were presented in \cite{mishali2010FromTheoryToPractice,Yonina_LPF,mishali751xampling,xampling_cit}. We now briefly review the sampling problem treated in \cite{mishali2010FromTheoryToPractice} and its relation to our setup. As we show the practical hardware implementation of both systems is similar.

\begin{figure}[h]
\centering
\includegraphics[scale=0.5]{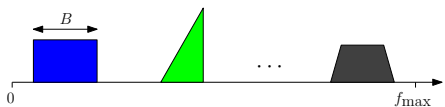}
\caption{Multiband signal model.}
\label{fig_multiband}
\end{figure}

The model in \cite{mishali2010FromTheoryToPractice} is of multiband signals: signals whose CTFT is concentrated on $N_{\textrm{bands}}$ frequency bands, and the width of each band is no greater than $B$. The location of the bands is unknown in advance. An example of such a signal is depicted in Fig.~\ref{fig_multiband}.
A low rate sampling scheme allowing recovery of such signals at a rate of $4 B N_{\textrm{bands}}$ was proposed in \cite{mishali2009blind}. This scheme exploits the sparsity of multiband signals in the frequency domain, to reduce the sampling rate well below the Nyquist rate.
In \cite{mishali2010FromTheoryToPractice,mishali751xampling,xampling_cit}, this approach was extended to a more practical sampling scheme, which uses a modulation stage and referred to as the Modulated Wideband Converter (MWC). In each channel of the MWC, the input is modulated with some periodic waveform, and then sampled using a LPF followed by a low rate uniform sampler.
The main idea is that in each channel, the spectrum of the signal is scrambled, such that a portion of the energy of all bands appears at baseband.
Mixing of the frequency bands in \cite{mishali2010FromTheoryToPractice} is analogous to mixing the Fourier coefficients in Fig.~\ref{fig_infinite_mixing_waveforms_sampling}.

We note here some differences between the methods.
First, following the mixing stage, we use an integrator in contrast to the LPF used in \cite{mishali2010FromTheoryToPractice}. This difference is a result of the different signal quantities measured: Fourier coefficients in our work as opposed to the frequency bands content in \cite{mishali2010FromTheoryToPractice}.
The second difference is in the purpose of the mixing procedure. In \cite{mishali2010FromTheoryToPractice} mixing is performed in order to reduce the sampling rate relative to the Nyquist rate. In our setting, the mixing is used in order to simplify the hardware implementation and to improve robustness to failure in one of the sampling channels.

Nonetheless, the hardware considerations in the mixing stage in both systems is similar. Recently, a prototype of the MWC has been implemented in hardware \cite{mishali751xampling}. This design is composed of $p=4$ sampling channels, where the repetition rate of the modulating waveforms is $1/T \approx 20\textrm{ MHz}$. In each period there are $N=108$ rectangular pulses.
This prototype, with certain modifications, can be used to implement our sampling scheme as well. These modifications mainly include adding shaping filters on modulating waveforms lines, and reducing the number of rectangular pulses in each period.


\section{Simulations}\label{sec_simulations}
In this section we provide several experiments in which we examine various aspects of our method.
The simulations are divided into $4$ parts:
\begin{enumerate}
\item Evaluation of the performance in the presence of noise, and comparison to other techniques;
\item Demonstration of the recovery method for pulse streams with SI structure;
\item Evaluation of the effects of synchronization errors between the channels;
\item Examination of the use of practical shaping filters.
\end{enumerate}

\subsection{Performance in the Presence of Noise}
We demonstrate the performance of our approach in the presence of white gaussian noise, when working at the rate of innovation.
We compare our results to those achieved by the integrators \cite{kusuma2006multichannel} and exponential filters \cite{Olkkonen} based methods, since these are the only approaches which can work at the same rate, for infinite stream of pulses.

We examine three modulation waveforms presented in Sections \ref{subsec_finite_mixing} and \ref{subsec_RelatedWork_SoS}: cosine and sine waveform (tones), filtered rectangular alternating pulses (rectangular) and waveforms obtained from delayed versions of the SoS filter (SoS).
For the rectangular pulses scheme, the modulation waveforms are generated using a single generator, as discussed in Section~\ref{subsec_pulses_waveform}. The shaping filter $g(t)$ is an ideal LPF with transition band of width $2\pi/T$.
Following \cite{Olkkonen}, the parameters defining the impulse response of the exponential filters are chosen as $\alpha=0.2T$ and $\beta=0.8T$.

\begin{figure}[h]
\centering
\includegraphics[]{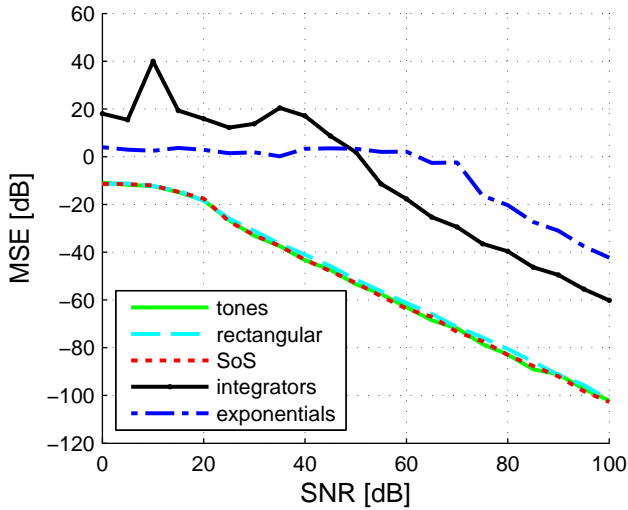}
\caption{Performance in the presence of noise, at the rate of innovation. The signal consists of $L=2$ Diracs.}
\label{fig_goyal_far}
\end{figure}

We focus on one period of the input signal, which consists of $L=2$ Diracs with $\bbf(t)=[0.256T,0.38T]^T$, and amplitudes $\bbf(a)=[1,0.8]^T$. We set the system parameters as $p=K=N=5$.
The estimation error of the time-delays versus the SNR is depicted in Fig.~\ref{fig_goyal_far}, for the various approaches. Evidently, our technique outperforms the integrators and exponential filters based methods in terms of noise robustness, for all configurations. There is a slight advantage of $2$dB for the schemes based on tones and SoS, over alternating pulses, where the first two configurations have similar performance.

\begin{figure}[h]
\centering
\includegraphics[]{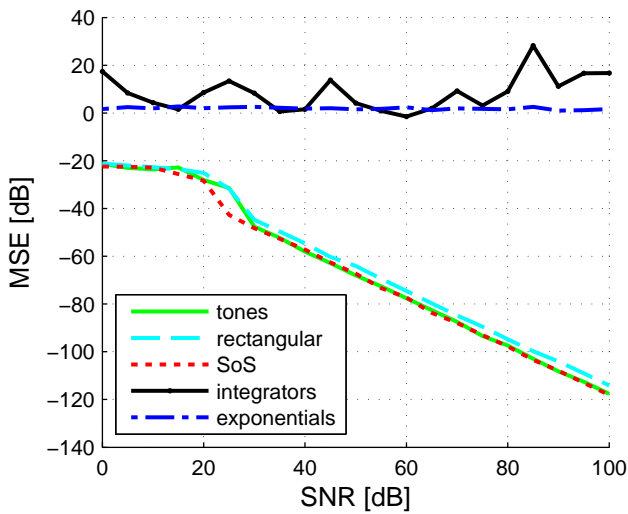}
\caption{Performance in the presence of noise, at the rate of innovation. The signal consists of $L=10$ Diracs.}
\label{fig_goyal_L10}
\end{figure}

Turning to higher order problems, in Fig.~\ref{fig_goyal_L10} we show the results for $L=10$ Diracs with times
chosen in the interval $[0,T)$ and amplitudes equal one, with $N=p=K=21$. The instability of the integrators and exponential filters based methods becomes apparent in this simulation. Our approach in contrast achieves good estimation results, demonstrating that our method is stable even for high model orders.

The performance advantage of the tones and SoS based schemes is now around $3.5$dB.
We conclude that from a noise robustness point of view, using multiple frequency sources or SoS waveforms is preferable over a single pulse generator. However, as discussed in Section~\ref{subsec_finite_mixing}, pulse sequences based schemes can be advantageous from practical implementation considerations, and reduce the hardware complexity. In addition, the performance degradation is reasonable, and the estimation error is still significantly lower than that of competing approaches.
Therefore, the flexibility of our architecture, allows the system designer to decide between better performance in the presence of noise, or lower hardware complexity.

Next, we compare our scheme to the one presented in \cite{DragottiStrangFix2007}. This approach, which is based on B-splines \cite{unser1999splines} and E-splines \cite{unser2005cardinal} sampling kernels, operates at a rate higher than the rate of innovation.
According to the main theorem in \cite{DragottiStrangFix2007}, an infinite stream of Diracs is uniquely determined from uniform samples taken at the output of a B-spline or E-spline sampling kernel, if there are at most
$L$ Diracs in an interval of size $2LSt_s$. Here $t_s$ is the sampling interval, and $S=2L$ is the time support of the sampling kernel.
In our setting there are $L$ Diracs in an interval of size $T$ requiring at least $p=(2L)^2$ samples per period $T$.

\begin{figure}[h]
\centering
\includegraphics[]{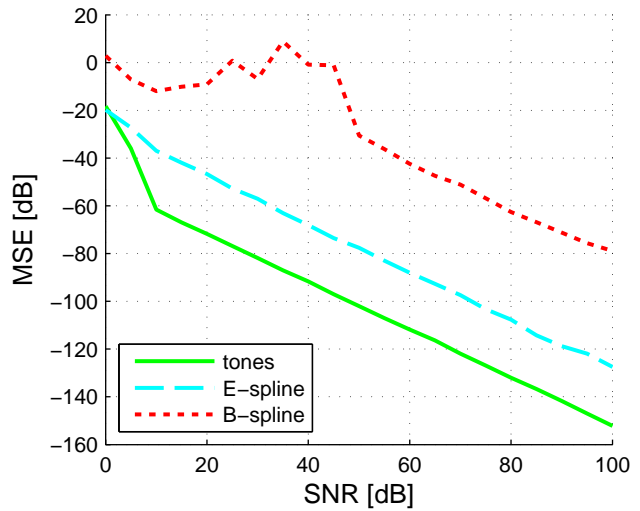}
\caption{Performance in the presence of noise, using $p=64$ samples. The signal consists of $L=4$ Diracs.}
\label{fig_drag_L4}
\end{figure}

We choose $L=4$ Diracs, with delays $\bbf(t)=[0.213T,0.452T,0.664T,0.745T]^T$ and amplitudes $\bbf(a)=[1,0.9,0.7,0.6]^T$.
We compare our tones based configuration to the both B-spline and E-spline techniques of \cite{DragottiStrangFix2007}. The parameters defining the E-spline kernel \cite{unser2005cardinal,DragottiStrangFix2007} were chosen in order to obtain real valued sampling kernels, and were tuned empirically to obtain the best performance.
For all algorithms $p=64$ samples are used. In order to exploit the oversampling in our approach, the Kumaresan and Tufts method \cite{kumaresan1983eaa} is used for the delays recovery.
The estimation error of the time-delays versus SNR is depicted in Fig.~\ref{fig_drag_L4}. From the figure it can be seen that our scheme exhibits better noise robustness than both B-spline and E-spline based methods.

\subsection{Sampling of Pulses with SI Structure}
\label{subsec_SI_sim}
We now consider sampling of streams of Diracs with SI structure.
We compare our method to the one presented in \cite{KfirYonina2009}. For our scheme, we examine two recovery options. The first is to process each period separately, namely, to recover the delays from each period independently (standard recovery). The second is to follow the approach presented in Section~\ref{subsec_infinite_SI} and to recover the common delays from all periods using the ESPRIT \cite{ESPRIT_Kailath} algorithm (SI recovery).

We consider $25$ periods with $L=4$ Diracs per period, and relative delays of $\bbf(t)=[0.213T,0.452T,0.664T,0.745T]^T$. The amplitudes in each period were taken as an independent Gaussian random variables, with means
$\boldsymbol{\mu}_a=[1,0.9,0.7,0.6]^T$ and standard deviation $\sigma=0.1$
For the method in \cite{KfirYonina2009}, we chose a single channel scheme with an ideal LPF as sampling kernel. For both methods $p=8$ samples per period were taken. The technique in \cite{KfirYonina2009} requires theoretically infinite number of samples, due to the infinite time support of the sampling kernel. However, to compare between the two methods, we used only $8 \cdot 25$ samples ($25$ periods, with $8$ samples per period).

\begin{figure}[h]
\centering
\includegraphics[]{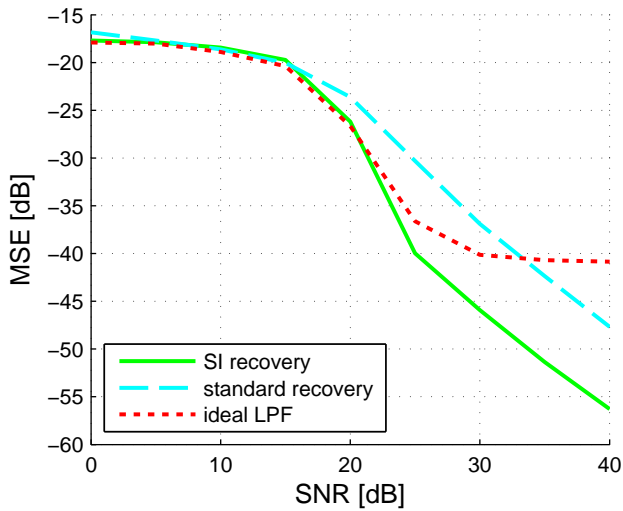}
\caption{Sampling of pulses with SI structure. The signal consists of $25$ periods, each with $L=4$ Diracs.}
\label{fig_SNR_SI}
\end{figure}

The estimation error of the time-delays versus the SNR is depicted in Fig.~\ref{fig_SNR_SI}.
For SNR levels above $15$dB, there is a clear advantage to the SI recovery method, over the standard approach. Hence, as expected, the use of the mutual information between periods on the delays, improves the estimation significantly.
Comparing the performance of our approach to the one of \cite{KfirYonina2009}, it can be seen that up to SNR levels of $20$dB both methods achieve similar performance. However, for higher SNRs, the method in \cite{KfirYonina2009} suffers from a dominant error caused by the fact that only a finite number of samples were used.
This demonstrates the advantage of our scheme, which operates on finite time intervals, in cases where the signal consists of a finite number of periods.

\subsection{Synchronization Errors}
\label{subsec_sync_sim}
We now study the error caused by synchronization errors between the channels.
We consider sampling of $L=4$ Diracs, with delays $\bbf(t)=[0.213T,0.452T,0.664T,0.745T]^T$ and amplitudes
$\bbf(a)=[1 0.9 0.7 0.6]^T$, using $p=9$ sampling channels.
We set the sampling time of the last channel to be shifted by $\Delta_{\textrm{max}}$ relative to the first channel. The offsets of the other channels are drawn uniformly on the interval $[0,\Delta_{\textrm{max}}]$.
We plot the standard deviation of the estimated time delays error, as a function of the maximal offset $\Delta_{\textrm{max}}$, for different SNR values.
We note that in this experiment, the misalignment between the sampling channels is not being compensated using the approach discussed in Section~\ref{subsec:sync}.

\begin{figure}[h]
\centering
\includegraphics[]{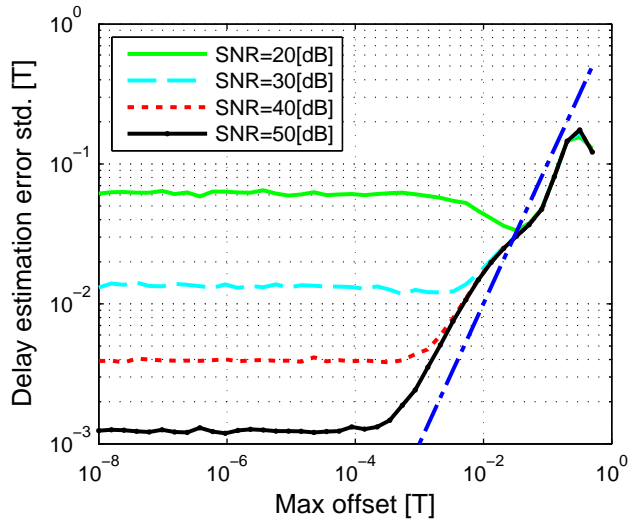}
\caption{Standard deviation of the time-delays estimation error, as a function of the maximal offset between channels, for various SNR levels.}
\label{fig_SNR_sync}
\end{figure}

The results are shown in Fig.~\ref{fig_SNR_sync}. The dash-dotted line denotes the linear curve $y=x$.
First, it can be seen that when the synchronization error is less than $10$ percent of the estimation error, the synchronization error is negligible, and the error is mainly due to the noise.
When the synchronization error becomes large, the time delay estimation error degrades linearly.
In general, the estimation error is bounded from below by the synchronization error (for $\Delta_{\textrm{max}}<0.03T$).

\subsection{Practical Shaping Filters}
\begin{figure}[h]
\centering
\includegraphics[]{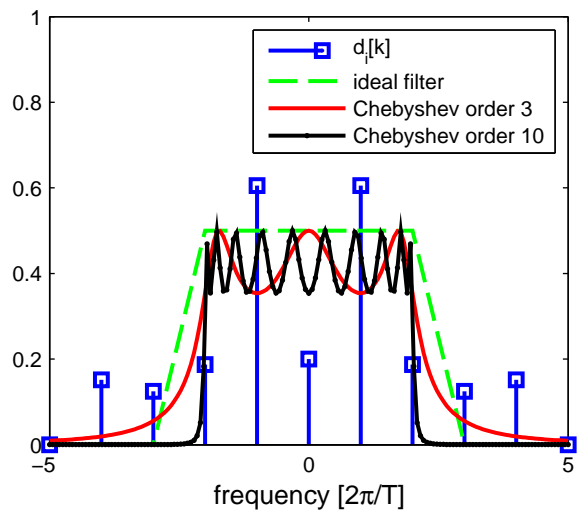}
\caption{Frequency responses of the shaping filters: ideal shaping filter vs. practical Chebyshev filters.}
\label{fig_cheby_freq}
\end{figure}

We now explore the use of practical shaping filters, for the rectangular pulses scheme, rather than the ideal ones used above. Once practical filters are used, the rejection of coefficients whose index is not in the set $\mathcal{K}$ is not perfect.
We set the shaping filter $g(t)$ to be a Chebyshev (Type I) LPF \cite{lam1979analog} of various orders, with ripple $3\textrm{ dB}$. The Chebyshev filter is a good choice for our requirements since it has a steeper roll-off than other filters, resulting in better rejection of the undesired coefficients. The rapid transition between the pass-band and stop-band of the Chebyshev filter comes at the expense of larger ripple in the pass-band, however, ripple is of minor concern for our method since it is digitally corrected when inverting the matrix $\bbf(S)$. The cutoff frequency was set to $\frac{2\pi}{T} \lfloor K/2 \rfloor$.
The frequency response of the various filters is shown in Fig.~\ref{fig_cheby_freq}. 

\begin{figure}[h]
\centering
\includegraphics[]{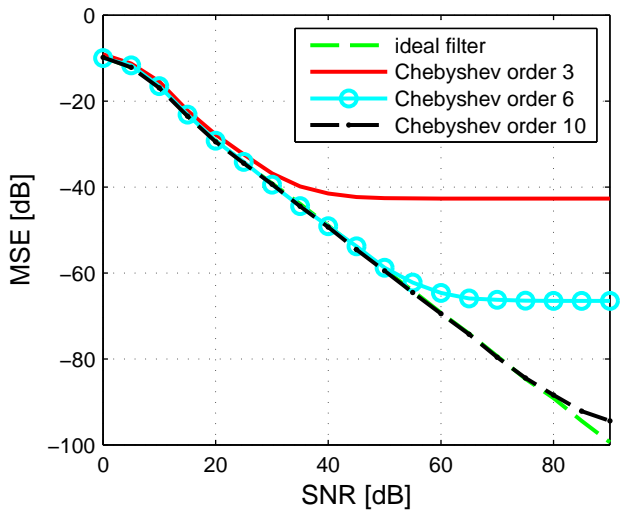}
\caption{Performance of practical shaping filters of various orders vs. ideal filtering.}
\label{fig_var_order}
\end{figure}

The estimation error of the time-delays versus the SNR is depicted in Fig.~\ref{fig_var_order}, for various filter orders. The simulation consists of $L=2$ Diracs with $\bbf(t)=[0.256T,0.46T]^T$, and amplitudes $\bbf(a)=[1,0.8]^T$.
Clearly, a Chebyshev filter of order $10$ closely approaches the performance of an ideal LPF. In addition, for SNR levels below $50$dB, using a Chebyshev filter of order $6$ provides good approximation.
Therefore, the modulation waveform generation stage of our proposed method can be implemented using practical analog filters.


\section{Conclusion}
In this work, we proposed a new class of sampling schemes for pulse streams. Our approach allows recovery of the delays and amplitudes defining such a signal, while operating at the rate of innovation.
In contrast to previous works \cite{kusuma2006multichannel,UnserFRI2008,Olkkonen} which achieved the rate of innovation, our approach supports general pulse shapes, rather than Diracs only.
In addition, as we demonstrate by simulations, our method exhibits better noise robustness than previous methods \cite{kusuma2006multichannel,Olkkonen,DragottiStrangFix2007}, and can accommodate high rates of innovation.

The proposed scheme is based on multiple channels, each comprised of mixing with a properly chosen waveform followed by an integrator. We exploit the degrees of freedom in the waveforms selection, and provide several useful configurations, which allow simplified hardware implementation and robustness to channel failure.
Using simulations we further explored practical issues, such as effects of misalignment between the sampling channels and usage of standard analog filters, in the waveform generation stage.

Our method can be viewed as a part of a broader framework for sub-Nyquist sampling of analog signals, referred to as Xampling \cite{mishali751xampling,xampling_cit}.
We draw connections with the work in \cite{mishali2010FromTheoryToPractice,mishali751xampling,xampling_cit}, which proposed a Xampling architecture for multiband signals. We showed that the hardware prototype of the analog front-end,
implemented for the multiband model, can be used in our scheme as well with certain modifications.

\section*{Acknowledgment}
The authors would like to thank the
anonymous reviewers for their valuable comments.

\bibliography{IEEEabrv,Multichannel_FRI_bib}

\end{document}